\documentclass[12pt,reqno,a4paper,twoside]{article}
\usepackage{graphicx}
\begin{document}

\title{{\bf
{Thinking transport as a twist}
}}

\author{
Cristian Giardin\`a
\footnote{Department of Mathematics and Computer Science,
Eindhoven University, P.O. Box 513 - 5600 MB Eindhoven, The Netherlands,
{\em c.giardina@tue.nl}}\\
Jorge Kurchan
\footnote{CNRS-ESPCI, rue Vauquelin 10, 75231 Paris, France,
{\em jorge@pmmh.espci.fr}}
}

\maketitle

\begin{abstract}
The   determination of the
conductivity of a  deterministic or stochastic classical system
coupled to reservoirs at its ends can  in general be mapped
 onto
the problem  of computing  the  stiffness (the `energy' cost of twisting the
boundaries) of a  quantum-like  system. The nature of the coupling to
the reservoirs determines the details of the mechanical coupling
of the torque at the ends.
\end{abstract}

\section{Introduction}

The transport properties of  physical systems reserve many surprises,
particularly in low dimensions, and despite many decades of efforts,
a general theory is surprisingly
not yet available \cite{BLR:00,LLP:03}.
This is true  even at the classical level.

The conductivity of low dimensional systems is often anomalous, with
transport coefficients diverging with the system size.  In such cases,
the nature of the contact with the reservoirs at the boundaries becomes an
issue. For example, the thermal conductivity of a finite chain becomes
zero not only in the obvious case in which the contact is bad, but
also in the limit when it is too good \cite{LLP:03,RLL:67}.
A conductivity computed on the basis of a closed chain
without reservoirs can clearly not take into account these effects, and
 in any
concrete physical realization of transport with anomalous properties,
one  has to consider the system as composed of both the bulk
and the bath.

Spin and charge currents in {\em closed quantum} chains  have been related to
the corresponding stiffnesses \cite{Kohn,Shastry}. This relation between
a transport and an equilibrium property has been very fruitful.
 More recently, the question of a quantum open chain without baths
has been addressed \cite{open}.

In this paper we follow a different path.  We shall establish a
general correspondence between {\em i)} the transport properties of any
classical (stochastic or deterministic) system {\em in contact with
  reservoirs} at its boundary,  and {\em ii)} the stiffness
(or helicity modulus \cite{fisher}) of a
zero-temperature  system that is perturbed at the ends by a
twist applied through two elastic `handles' (see Figure 1).  In this
mapping, the details of the coupling between these handles and the
system are important and reflect the nature of the coupling between
reservoirs and the lattice in the original setting.

As we shall see, the correspondence we discuss here works
at a different level from the one of
Kohn \cite{Kohn} and Shastry-Sutherland  \cite{Shastry}. Here, the stiffness
one calculates is that of the {\em evolution operator}~\cite{Gaspard} and
not of the energy itself~\footnote{
And in the quantum
 generalization of the present work,
the twist will act on the closed time path action, involving twice as
 many fields, and not on the original Hamiltonian.}.

The plan of the paper is the following.  In the next section we
introduce our setting, recall the definition of conductivity and of
stiffness and explain our strategy.  We will find convenient to use
the operator bra-ket notation by which the probability distribution of
Eq. (\ref{master}) is associated to a Hilbert space quantum-like
``state''; this is described in Section 3, together with two examples
that will be repeatedly used in the paper.  Armed with this
preliminaries, we first show in Section 4 that the bulk current in a
transport model subject to a gradient in the boundary conditions can
be expressed in terms of a `helical' operator. We then use linear
response theory to rederive in Section 5 two equivalent expressions of
the finite volume conductivity: the first coincides with the standard
Green-Kubo formula (i.e., time integral of the {\em bulk} current
autocorrelation function); the second one involves the autocorrelation
function of an operator which depends only on the {\em boundaries}.
This second expression allows us to show that the conductivity is
proportional to the stiffness of the evolution operator when a twist
is exerted at the boundaries. This is shown in Section 6, which is
then followed by conclusions.

\section{Transport model}

\begin{figure}
\begin{center}
\includegraphics[width=7.cm]{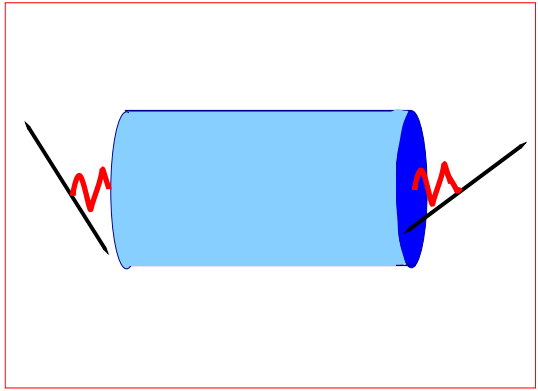}
\label{poincare}
\caption{{\small Stiffness of a bar twisted by  elastic handles. }}
\end{center}
\label{fig1}
\end{figure}

Consider an extended system made of $N$ components whose temporal
evolution is either deterministic (Hamiltonian) or stochastic and
whose boundaries are coupled to  reservoirs.  A convenient
description of the system is given by the evolution
(Fokker-Planck, Liouville,...)
equation
\begin{eqnarray}
\label{master}
\frac{\partial}{\partial t}\,\mu(x,t) &=& - H \mu(x,t)\;,
\nonumber \\
\mu(x,t) &=& e^{-tH} \mu(x,t=0)
\end{eqnarray}
where
\begin{itemize}
\item $\mu(x,t)$ is the probability distribution in a point $x$ of the
phase/configurations space at time $t$.
\item
$H$ is the evolution operator which is assumed of the form
\begin{equation}
 H=H_{B} + \lambda\Big(H_L(\alpha_L) + H_R(\alpha_R)\Big)\;.
\end{equation}
\end{itemize}
Here $H_B$ denotes the bulk evolution, whereas $H_{L}(\alpha_L)$ and
$H_{R}(\alpha_R)$, which depend on a parameter $\alpha$, describe the
interaction with reservoirs connected at the left/right boundaries,
respectively.  We have made  explicit the strength of the coupling
 to the reservoirs with the parameter $\lambda$.

\subsection{ Conductivity}
We consider situations when
 the {\em bulk} evolution has a constant of motion $E(x)$ -
typically the energy or the mass. We shall assume $E(x)$ is spatially
local: it is the sum of N local contributions
\begin{equation}
\label{conserved}
E= \sum_{i=1}^N E_i\;,
\end{equation}
each $E_i$ a function of the site $i$ or its near-neighbors.

A probability distribution function that is concentrated on an energy 
(or any other conserved quantity) shell $\delta[E(x)-E_o]$
satisfies:
\begin{equation}
 E(x) \mu(x,t) = E_o \mu(x,t)
\end{equation}
so that the Hilbert space breaks into subspaces corresponding 
to different values of the conserved quantity.
If the stochastic evolution $H_B$ is conservative, there are no transitions
between shells, so $H_B$ is of block form, each one corresponding
to a value of $E_o$.
This in turn implies that $E$ commutes with the bulk evolution operator:
\begin{equation}
\label{derivata-energia}
 [E, H_B] = 0\;.
\end{equation}
The reservoirs violate the conservation of  $E$
\begin{equation}
\label{derivata-energia1}
 [E, H_L] \neq 0  \;\;\;\;\;\; ; \;\;\;  \;\;\; [E, H_R] \neq 0\; ,
\end{equation}
and, in addition, if the   values of
the parameter $\alpha_L \neq \alpha_R$ are different, a current is established,
  a stationary state eventually sets in and one has
transport of the quantity $E$ from one boundary to the other. Examples
are:
\begin{itemize}
\item[-] when the parameter is the temperature, $\alpha = T$, one has transport
of energy;
\item[-] when the parameter is a particle density, $\alpha = \rho$, one has
transport of mass.
\end{itemize}

If $\alpha_R-\alpha_L$ is small, in the stationary regime the
averaged total current across all links $\langle{\cal J }\rangle$ is linear:
\begin{equation}
\langle{\cal J}\rangle \sim \kappa_N ( \alpha_R-\alpha_L)
\end{equation}
which defines the conductivity $\kappa_N$ for a system of size $N$.

\subsection{Stiffness}

In order to recall the definition of the stiffness let us consider
a simple example  totally unrelated to the context of  
transport models. Suppose we are given a {\em quantum} Hamiltonian
\begin{equation}
{\bar H } = {\bar H_B} + \lambda ({\bar H_L} + {\bar H_R})
\end{equation}
with a bulk part
\begin{equation}
{\bar H}_B= - \sum_{i=1}^N \hbar^2 \frac{\partial^2}{\partial x_i^2}
- \sum_{i=1}^{N-1}\cos(x_{i+1}-x_{i})
\end{equation}
which clearly has symmetry with respect to simultaneous shift of all angles.
The expectation of the global angle is fixed by two boundary terms:
\begin{equation}
{\bar H}_L=- \cos(x_1) \;\;\;\; ; \;\;\;\;
{\bar H}_R=-\cos(x_N)
\end{equation}
 which impose an optimal average profile $\langle x_i \rangle = 0$
 for all $i\in\{1,\dots,N\}$.
If we wish to calculate the stiffness, taking into account the specific
coupling, we do the following: we compute first the lowest eigenvalue
$\epsilon^o$ of ${\bar H}_B + \lambda({\bar H}_L + {\bar H}_R)$.
Next, we twist the two `handles'
\begin{equation}
{\bar H}_L^{{\mbox{\small twisted}}}= -\cos(x_1+\theta)
\;\;\; ; \;\;\;
{\bar H}_R^{{\mbox{\small twisted}}}= -\cos(x_N-\theta)
\end{equation}
and calculate the new lowest eigenvalue $\epsilon^{\theta}$.  In the
limit where the coupling $\lambda\to 0$ then, of course, imposing the
twist has no effect. In the opposite limit $\lambda\to \infty$ then
$x_1 = -\theta$ and $x_N = \theta$ are strictly imposed.  For a finite
value of the coupling $0< \lambda < \infty$ there is a competition
between the bulk lattice which prefers a flat profile and the
boundaries which force a gradient.  In the framework of elasticity
theory, the stiffness $\sigma_N$ measures the cost of twisting and
 is defined from
\begin{equation}
\epsilon^{\theta}-\epsilon^{o}
\sim \frac12 \frac{\sigma_N}{(N-1)} (2\theta)^2\;,
\label{iii}
\end{equation}
 valid to first order in $\theta^2$. The factor $N-1$ assures a good
scaling for large $N$ in a system with normal elasticity.

To define the stiffness of the transport model, which includes bulk
and reservoirs, we proceed in a similar manner.  We think of the
evolution operator $H$ in formula (\ref{master}) as a quantum-like,
albeit non-Hermitian, Hamiltonian. We can interpret
(\ref{derivata-energia}) as an invariance of the bulk with respect to
`rotations' of angle $\theta$ generated by a group $e^{i\theta E}$.

The boundary couplings break this invariance (cfr. (\ref{derivata-energia1})),
we can use them as `handles' to impose a `twist' by applying
the transformation in opposite directions at the ends:
\begin{equation}
H_R^{{\mbox{\small twisted}}}=  e^{i\theta E} H_R e^{-i\theta E}
\;\;\; ; \;\;\;
H_L^{{\mbox{\small twisted}}}=  e^{-i\theta E} H_L e^{i\theta E}
\label{twist}
\end{equation}
The (zero-temperature) stiffness $\sigma$ of the handle+bulk system
is defined as in (\ref{iii}) from the increase  in the lowest eigenvalue of
$H^{{\mbox{\small twisted}}}(\theta)=H_R^{{\mbox{\small twisted}}}+H_B+
H_L^{{\mbox{\small twisted}}}$ for small $\theta$.

The results we shall show in what follows is that {\em the conductivity
$\kappa$ is, up to a trivial factor, equal to the stiffness $\sigma$}.
In general, anomalous conductivity amounts to the system becoming {\em rigid},
due to long-range correlations.
When the coupling to the bath is weak, the conductivity is small: the
stiffness is also weak because  the handles turn
 without affecting the system. In the opposite case, when the coupling
is too strong, the conductivity of a system with anomalous diffusion
may go to zero.  From the point of view of elasticity, what happens is that
the handles twist  strongly the first and last sites of the chain, and
it is more favorable for a stiff system to concentrate
all the twist  between the first two (and between the last two)
sites.

\section{Bracket Notation}

A convenient way to write the evolution equation (\ref{master}) is
provided by the bra-ket formalism. The probability distribution
at time $t$ is encoded in the state $|\psi(t)\rangle$, namely
\begin{equation}
|\psi(t) \rangle = \int dx \mu(x,t) |x\rangle \;,
\end{equation}
where $|x\rangle$ denotes a vector which together with its transposed
$\langle x|$ form a complete basis of a Hilbert space and its dual, that is
\begin{equation}
\langle x|x' \rangle = \delta(x-x')\;.
\end{equation}
It immediately follows that
\begin{equation}
\langle x|\psi(t) \rangle = \mu(x,t)
\end{equation}
and the evolution equation (\ref{master}) takes the form of a
Schr\"odinger equation with imaginary time
\begin{equation}
\label{schroedinger}
\frac{d}{dt}|\psi(t) \rangle = -H |\psi(t)\rangle\;.
\end{equation}
To compute expectation we introduce the flat state
\begin{equation}
\langle - | = \int dx \;\langle x|
\end{equation}
which is such that
\begin{equation}
\langle - | x\rangle = 1 \;.
\end{equation}
Then for any observable $A$ we have that its expectation value
at time $t$ can be written as
\begin{equation}
\langle A(t) \rangle = \int dx \mu(x,t) A(x) = \langle -|A|  \psi(t) \rangle \;.
\end{equation}
Conservation of probability implies that
\begin{equation}
\langle -|H=0 \;.
\end{equation}
The stationary state $|\psi\rangle$ satisfies
\begin{equation}
H |\psi \rangle = 0\;.
\end{equation}
For an isolated system, namely $\lambda = 0$, the invariant measure are
given by any arbitrary function of $E$, as it is immediately seen from
Eq.(\ref{derivata-energia}). For example the microcanonical ensemble
is given by the uniform measure
\begin{equation}
|\psi\rangle_{e} =  \int dx \;\delta(E(x)- Ne)|x\rangle\;.
\end{equation}
In the presence of reservoirs, namely $\lambda \neq 0$,
the stationary state is in general not known. The boundaries
operators $H_{L/R}$ representing the action of the reservoirs are chosen
such that, when the bath parameters are equal $\alpha_L =\alpha_R = \alpha$
the unique invariant measure is given by the equilibrium Boltzman-Gibbs
measure
\begin{equation}
\label{bg}
|\psi\rangle_{\tilde\alpha} = \int dx \;\frac{e^{-\tilde\alpha
    E(x)}}{Z}|x\rangle\;,
\end{equation}
where $Z$ is the normalizing partition function

In the following we will denote the equilibrium Boltzman-Gibbs state
$|\psi\rangle_{\tilde\alpha}$ with
$|\tilde\alpha\rangle$.
In case the system has a discrete configuration space, integrals
in all the previous formulas are replaced by discrete sums and
Dirac delta functions are replaced by Kronecker delta functions.

The relation between $\alpha$ and $\tilde \alpha$ is easy to obtain
for each problem using the fact that
the equilibrium state $|\tilde\alpha\rangle$,
see Eq. (\ref{bg}), is annihilated by the boundary `bath'
operators $H_{L/R}(\alpha)$
\begin{equation}
\label{333}
\Big( H_{L/R} (\alpha) \Big) |  \tilde\alpha \rangle =0\;,
\end{equation}
This is because, by assumption, the bath terms leave the corresponding
equilibrium structure stationary.

\vspace{1.cm}

{\bf Example A:} {\em Hamiltonian systems}

A large class of systems is represented by a model
composed of $N$ point-like particles interacting with their
nearest neighbours. This is described by the Hamiltonian
\begin{equation}
\label{hamiltonian}
E = \sum_{i=1}^N \frac{ p_i^2}{2} + \sum_{i=1}^{N-1} V(q_{i+1}-q_i)\;.
\end{equation}
The evolution equation (\ref{master}) holds with
$x=(q_1,\ldots,q_N,p_1,\ldots,p_N)$ a point in phase space. The bulk
evolution operator is the Liouville operator
\begin{equation}
H_B= \{\;,E\} = \sum_{i=1}^N p_i\frac{\partial}{\partial q_i} -
\frac{\partial V}{\partial q_i}\frac{\partial}{\partial p_i}\;,
\end{equation}
where $\{\;,\;\}$ denotes the Poisson brackets, that is for any
functions $f$ and $g$
\begin{equation}
\{f,g\} = \sum_{i=1}^N \frac{\partial f}{\partial q_i }\frac{\partial
  g}{\partial p_i } - \frac{\partial f }{\partial p_i }\frac{\partial
  g }{\partial q_i } \;.
\end{equation}
The constant of motion for the bulk evolution is the
total bulk energy $E$. Indeed $H_B$ obviously
commutes with the operator that multiplies by $E$:
\begin{equation}
H_B (E f)= \{E f,E\}= E \{f,E\} + f\{E,E\} = E \{f,E\} = E (H_B f) \qquad
\forall f\;.
\end{equation}
To model the interaction with the boundaries, stochastic heat
reservoirs can be taken as Ornstein-Uhlenbeck processes whose
variances $T_{L/R}$ specify the temperature of the bath. This is
represented by boundaries evolution operators
\begin{equation}
\label{bathL}
- H_L(T_L)=\frac{\partial}{\partial
  p_1}\left(T_L\frac{\partial}{\partial p_1}+p_1\right)\;,
\end{equation}
\begin{equation}
\label{bathR}
- H_R(T_R)=\frac{\partial}{\partial
  p_N}\left(T_R\frac{\partial}{\partial p_N}+p_N\right)\;.
\end{equation}
When the two thermal reservoirs have the same temperature $T_L = T_R = T$,
the stationary state $|\psi\rangle_{\beta}$
is the Boltzman-Gibbs equilibrium measure
\begin{equation}
|\psi\rangle_{\beta} = |\beta \rangle = \int dx \frac{e^{-\beta
    E(x)}}{Z}|x\rangle\;,
\end{equation}
where the inverse temperature $\beta = 1/T$ is such that
\begin{equation}
\langle E \rangle = \langle -| E | \beta \rangle = T\;.
\end{equation}

\vspace{1.cm}

{\bf Example B:} {\em Simple symmetric exclusion process}

Another class of models we consider are continuous time Markov
processes. With the restriction of a finite configuration space ${\cal
  S}$, the process is specified by assigning the rates of transition
$w(\sigma',\sigma)$ for jumping from a configuration $\sigma\in {\cal
  S}$ to a configuration $\sigma'\in {\cal S}$.  The master equation
for $\mu(\sigma,t)$, the probability distribution of a configuration
$\sigma$ at time $t$, then reads
\begin{equation}
\frac{d\mu(\sigma,t)}{dt} =
\sum_{\sigma'\neq\sigma}
\left(
w(\sigma,\sigma')\mu(\sigma',t) - w(\sigma',\sigma)\mu(\sigma,t)
\right)
= - H \mu(\sigma,t)\;.
\end{equation}
To be more definite we will work with the simple symmetric exclusion
process which can be expressed - in a operator formalism -- as an
 SU(2) ferromagnet of spin $1/2$ \cite{Schutzsandow}.
This corresponds to the stochastic process on the lattice $\{1,\ldots,N\}$
where particles jumps at rate 1 to one of their neighbours and each site
can accomodate at most $1$ particle per site.
Configurations $n\in\{0,1\}^N$ are then identified with ket states
\begin{equation}
|n\rangle = |n_1,\ldots,n_N\rangle = \otimes_{i=1}^N |n_i\rangle\;,
\end{equation}
which specify the occupation numbers of each sites, namely
$n_i\in\{0,1\}$.  The bulk evolution is given by
 the transition rates
\begin{eqnarray}
w(n^{i+1,i},n) & = & - \langle n^{i,i+1}|H_B|n\rangle
=(1 - n_i)n_{i+1}\nonumber \\ w(n^{i,i+1},n) & = & - \langle
n^{i,i+1}|H_B|n\rangle =n_i(1 - n_{i+1})\;.
\end{eqnarray}
where $n^{i,j}$ is the configuration which is obtained from the
configuration $n$ by removing a particle in $i$ and adding it in $j$.

In operator notation, this is generated by \cite{Schutzsandow}
\begin{equation}
\label{bulksu2}
- H_B = \sum_{i=1}^{N-1} \left( S^{+}_iS^{-}_{i+1} +
S^{-}_iS^{+}_{i+1} + 2 S^{0}_iS^{0}_{i+1} - \frac12 \right)\;,
\end{equation}
where the $S$'s operators act as
\begin{eqnarray}
S^+_i |n_i\rangle &=& (1-n_i) |n_i+1\rangle \nonumber \\
S^-_i |n_i\rangle &=& n_i      |n_i-1\rangle \nonumber \\
S^0_i |n_i\rangle &=& \left(n_i-\frac12\right)  |n_i\rangle\;.
\end{eqnarray}
and
satisfy the SU(2) algebra
\begin{eqnarray}
\label{commutatorsSU2}
[S_i^{0},S_i^{\pm}] &=& \pm S_i^{\pm} \nonumber \\
{[}S_{i}^{-},S_{i}^{+}{]} &=& -2S_i^{0}\;.
\end{eqnarray}

In this stochastic system the constant of motion - in the absence of
reservoirs - is the total number of particles
\begin{equation}
\label{Esep}
E = \sum_{i = 1}^{N} \left(S^{0}_i + \frac12 \right)\;.
\end{equation}
This is obviously a conserved quantity for a particle jump process,
\begin{equation}
[E,H_B] =0\;,
\end{equation}
as can be immediately checked using the commutation relations
(\ref{commutatorsSU2}).

When the system is coupled to reservoirs at different chemical
potential, the boundaries can inject or absorb particle from the
system.  We assume that particles are injected at site $1$ at rate
$\rho_L$ and they are removed from site $1$ at rate $(1-\rho_L)$;
in the same way particles are injected at site $N$ at rate
$\rho_R)$ and they are removed from site $N$ at rate $(1-\rho_R)$.
The boundaries evolution operators then read
\begin{equation}
\label{reservL}
- H_L(\rho_L) = \rho_L \left(S_1^{+} + S_1^{0} - \frac12 \right) +
(1-\rho_L)\left(S_1^{-} - S_1^{0} - \frac12 \right)
\end{equation}
\begin{equation}
\label{reservR}
- H_R(\rho_R)= \rho_R\left(S_N^{+} + S_N^{0} - \frac12 \right) +
(1-\rho_R)\left(S_N^{-} - S_N^{0} - \frac12 \right)
\end{equation}
The constants $\rho_L$ and $\rho_R$ are to interpreted as the
two densities the reservoirs would impose if acting separately.
When the two reservoirs have the same densities $\rho_L = \rho_R = \rho$,
the stationary state $|\psi\rangle_\nu$
is the Boltzman-Gibbs equilibrium measure
\begin{equation}
|\psi\rangle_\nu = |\nu \rangle = \sum_{\{n\}} \frac{e^{\nu E}}{Z}|n\rangle\;,
\end{equation}
where the chemical potential $\nu = \ln
\left(\frac{\rho}{1-\rho}\right)$ is such that
\begin{equation}
\langle E \rangle = \langle -| E | \nu \rangle = \rho\;.
\end{equation}
As it is immediately checked, in this case the equilibrium Gibbs
measure is a Bernoulli product measure with parameter $\rho =
e^{\nu}/(1+e^{\nu})$
\begin{equation}
\label{sep-eq}
|\rho\rangle = \otimes_{i=1}^N |\rho\rangle_i = \otimes_{i=1}^N
\Big(\rho\, |1\rangle_i + (1-\rho)\,|0\rangle_i\Big)\;.
\end{equation}

\section{Bulk current and `helical' operator}

We wish to study the transport of the quantity $E$ in the presence of
reservoirs. Our first step is to express the current in the bulk
in terms of a `helical' operator.

The current which passes through the system is defined via
the continuity equation
\begin{equation}
\frac{d E}{d t} + \nabla {\cal J} = 0\;.
\end{equation}
Since $E$ is the sum of $N$ local contributions, see
Eq.(\ref{conserved}), we can write a local continuity equation as
\begin{equation}
\label{continuitylocal}
\frac{d E_s}{d t} =  - ({\cal J}_{s,s+1} - {\cal J}_{s-1,s})\;,
\qquad\qquad s = 1,\ldots,N\;.
\end{equation}
where ${\cal J}_{s,s+1}$ is the current from site $s$ to site $s+1$,
and  the total {\em bulk} current is defined as
\begin{equation}
{\cal J} = \sum_{s=1}^{N-1} {\cal J}_{s,s+1}\;.
\end{equation}

For a stochastic system, $ {\cal J}_{s,s+1}$ may depend explicitely on the
random noise. We shall consider a set of  operators $J_{s,s+1}$ such that
they coincide  with the average of ${\cal J}_{s,s+1}$
over random  realisations
\begin{equation}
\langle  {\cal J}_{s,s+1} (t)\rangle_{{\mbox{\tiny noise}}} =
\langle -|J_{s,s+1} |\Psi(t) \rangle \;\;\;\;\; ; \;\;\;\;\;
 \langle  {\cal J} (t)\rangle_{{\mbox{\tiny noise}}} =
\langle -|J |\Psi(t) \rangle
\end{equation}

Defining the  {\em helical} operator
\begin{equation}
A= \sum_{s=1}^N s E_s\;.
\end{equation}
 $J$ can be
expressed as the commutator between it and
the bulk evolution operator, that is
\begin{equation}
\label{maincommutator}
 J = [A,H_B]\;
\end{equation}
Indeed, for an open chain (namely $\lambda = 0$) we have
\begin{equation}
\langle -| \left[ [A,H_B]-\frac{dA}{dt}\right] | \Psi(t) \rangle=0\;.
\end{equation}
Making use of the continuity equation (\ref{continuitylocal}), we then have
\begin{equation}
\langle -|[A,H_B]| \Psi(t) \rangle=
\langle -|\frac{dA}{dt}| \Psi(t) \rangle=
\sum_{s=1}^N s \langle -|\frac{dE_s}{dt}| \Psi(t) \rangle =
\sum_{s=1}^N s \langle  {\cal J}_{s-1,s}-{\cal J}_{s,s+1} \rangle\;.
\end{equation}
Since ${\cal J}_{0,1}={\cal J}_{N,N+1}=0$ for an open chain:
\begin{equation}
\langle -|[A,H_B]| \Psi(t) \rangle =
 \sum_{s=2}^N s\langle -| { J}_{s-1,s}| \Psi(t) \rangle
 - \sum_{s=1}^{N-1} s \langle -|{J}_{s,s+1}| \Psi(t) \rangle\;,
\end{equation}
and making the change $s\to s+1$ in the first summation we find
\begin{equation}
\langle -|[A,H_B]|\Psi(t) \rangle=
 \sum_{s=1}^{N-1} \langle {\cal J}_{s,s+1}\rangle  = \langle {\cal J}\rangle =
 \langle -|J| \Psi(t) \rangle\;.
\end{equation}

{\bf Remark 1:} {\em The correspondence between $J$ and ${\cal J}$ is via
expectation values. Note however that expectations of products yield extra
terms in stochastic systems:
$\langle {\cal J}^2\rangle =\langle -|J^2 |\Psi \rangle
+ {\mbox{ extra trems}}$.
This is the origin of extra terms in the formulas below.}

{\bf Remark: 2} {\em In the definition of the helical operator
one can always add a term whose total derivative
with respect to time is zero. Later, to make the role of the
boundaries more symmetric, we will consider}
\begin{equation}
A = \sum_{s=1}^N s E_s - \frac{N+1}{2} E
\end{equation}

\vspace{0.5cm}
{\bf Example A:} {\em Hamiltonian system}

For the system described by the Hamiltonian (\ref{hamiltonian})
the current from site $s$ to site $s+1$ is given by
(see \cite{LLP:03})
\begin{equation}
{\cal J}_{s,s+1} = - \frac12 (p_s + p_{s+1}) V'(q_{s+1}-q_s)
\end{equation}
Defining the local operators $E_s$ as
\begin{eqnarray}
E_1 & = & \frac{p_1^2}{2} + \frac12 V(q_2-q_1) \nonumber \\ E_s & = &
\frac{p_s^2}{2} + \frac12 \Big(V(q_{s+1}-q_{s})+V(q_{s}-q_{s-1})\Big)
\qquad\qquad s = 2,\ldots,N-1 \nonumber \\ E_N & = & \frac{p_N^2}{2} +
\frac12 V(q_{N}-q_{N-1})
\end{eqnarray}
one can check that (\ref{maincommutator}) holds with
\begin{eqnarray}
A & = & \sum_{s=1}^N s E_s - \frac{N+1}{2} E \nonumber \\ & = &
\sum_{s=1}^N s\frac{p_s^2}{2} + \sum_{s=1}^{N-1}(2s+1) V(q_{s+1}-q_s)
- \frac{N+1}{2} E
\end{eqnarray}

\vspace{0.5cm}
{\bf Example B:} {\em Simple symmetric exclusion process}

For the system described by the evolution
operator (\ref{bulksu2})
the current from site $s$ to site $s+1$ is
\begin{equation}
{J}_{s,s+1} = S_s^{-}S_{s+1}^{+} - S_s^{+}S_{s+1}^{-}
\end{equation}
Equation  (\ref{maincommutator}) now holds with
\begin{eqnarray}
A & = & \sum_{s=1}^N s E_s - \frac{N+1}{2} E
\nonumber \\
& = &
\sum_{s=1}^L s \left(S^{0}_s + \frac12 \right) - \frac{N+1}{2} E
\end{eqnarray}

\subsection{A useful general identity}

The boundary terms $H_{L/R}(\alpha)$ break the conservation law for
the quantity $E$. They depend on a parameter $\alpha$ (e.g. the temperature
in the Hamiltonian system example, the chemical potential in the
symmetric exclusion process example) and impose a unique stationary
equilibrium Boltzmann Gibbs state $|\psi\rangle_{\tilde\alpha} =
|\tilde\alpha\rangle$.  We wish to find a convenient way to express
the response of the boundary operators $H_{L/R}(\alpha)$ when their
parameter $\alpha$ is varied. We claim the following identities hold:
\begin{equation}
\label{bathresponse}
\left( \frac{\partial}{\partial \alpha} H_{L/R} (\alpha)\right)
|\tilde\alpha \rangle = c_{\alpha}\; [H_{L/R}(\alpha),E] \;
|\tilde\alpha \rangle\;,
\end{equation}
where the constant $c_{\alpha}$ is given by
\begin{equation}
c_{\alpha} = \frac{\partial \tilde\alpha}{\partial \alpha}\;.
\end{equation}

This can be proved as follows.
We use the fact that the  Gibbs-Boltzmann distribution is annihilated
by the boundary  operators (cfr Eq. (\ref{333})),
and that, because of the Gibbs-Boltzmann form,  it satisfies:
\begin{eqnarray}
\label{derivbeta}
\frac{\partial}{\partial \tilde\alpha} |\tilde\alpha \rangle &=&
-(E-\langle E \rangle) \;   |\tilde\alpha \rangle
\label{deriv}
\end{eqnarray}
where $\langle E \rangle=\langle -|E|  \tilde\alpha \rangle$.
Computing the derivative of Eq.(\ref{333}) with respect to $\tilde\alpha$,
this in turn gives:
\begin{eqnarray}
0 &=&\left( \frac{\partial}{\partial \tilde\alpha} H_{L/R}
(\alpha)\right) | \tilde\alpha \rangle + H_{L/R} (\alpha)
\frac{\partial}{\partial \tilde\alpha} |\tilde\alpha \rangle \nonumber
\\ &=&\left( \frac{\partial}{\partial \tilde\alpha} H_{L/R}
(\alpha)\right) | \tilde\alpha \rangle - H_{L/R} (\alpha) (E-\langle E
\rangle) \; |\tilde\alpha \rangle \nonumber \\ &=&\left(
\frac{\partial}{\partial \tilde\alpha} H_{L/R} (\alpha)\right) |
\tilde\alpha \rangle - [H_{L/R}(\alpha),E] \; |\tilde\alpha \rangle
\label{uno}
\end{eqnarray}
where we have used (\ref{derivbeta}). Eq. (\ref{bathresponse}) then follows.

\vspace{0.5cm} {\bf Example A:} {\em Hamiltonian system}

Eq. (\ref{bathresponse}) holds with $\alpha = T$, $\tilde\alpha =
\beta = \frac{1}{T}$ and with $c_{T} = - \frac{1}{T^2}$.

\vspace{0.5cm}
{\bf Example B:} {\em Simple symmetric exclusion process}

Eq. (\ref{bathresponse}) holds with $\alpha = \rho$, $\tilde\alpha =
-\nu = \ln\left(\frac{1-\rho}{\rho}\right)$ and with $c_{\rho} =
-\frac{1}{\rho(1-\rho)}$.

\section{Linear response theory}

The conductivity is obtained by calculating the average current in the
presence of a small mismatch in the parameters of the reservoirs
evolution operators.  If the left reservoir is working with a
parameter $\alpha_L = \alpha - \delta \alpha$ and the right reservoir
is working with a parameter $\alpha_R = \alpha + \delta \alpha$, the
evolution  equation
\begin{equation}
\frac{d}{dt}|\psi(t) \rangle = -H |\psi(t)\rangle
\end{equation}
can be solved in a linear response regime.
Expanding to first order in $\delta \alpha$
we find
\begin{equation}
H = H_0 + \delta \alpha \;\lambda \;\Delta H + o(\delta\alpha)
\end{equation}
with
\begin{equation}
H_0 = H_B + \lambda (H_R(\alpha) + H_{L}(\alpha))
\end{equation}
\begin{equation}
\Delta H = \left(  \frac{\partial}{\partial \alpha} H_R
 -\frac{\partial}{\partial \alpha} H_L \right)
\end{equation}
and
\begin{equation}
|\psi(t)\rangle = |\tilde\alpha\rangle + \delta\alpha \;
|\Delta\psi(t)\rangle + o(\delta\alpha)
\end{equation}
with $|\tilde \alpha\rangle$ is the stationary state for
the unperturbed problem, that is
\begin{equation}
\frac{d}{dt}|\tilde \alpha \rangle = -H_0 |\tilde \alpha\rangle = 0
\end{equation}
The solution is
\begin{equation}
|\Delta \psi(t)\rangle = - \lambda \int_0^t dt' \; e^{-(t-t')H} \Delta
H |\tilde \alpha \rangle
\end{equation}
To leading order the average value of the total bulk current will be
\begin{eqnarray}
\langle {\cal J}(t)\rangle
& = & \langle -| { J} |\psi(t)\rangle
\nonumber \\
& = &
\langle -| {J} | \tilde \alpha\rangle \,-\,\delta\alpha \,\lambda
\int_0^t dt' \;
\langle -| {J} e^{-(t-t')H} \Delta H |\tilde \alpha \rangle
+ o(\delta\alpha)
\label{cosa2}
\end{eqnarray}
The conductivity $\kappa_N$ for a system of size $N$ is defined as
\begin{equation}
\kappa_N = \lim_{t\rightarrow\infty}
\,\frac{\delta \langle {\cal J}(t)\rangle}{2\,\delta \alpha}
\end{equation}
From Eq.(\ref{cosa2}) the following formula for the conductivity is
then deduced:
\begin{equation}
\label{gk1}
\kappa_N =  \lim_{t\rightarrow\infty} \,-
\frac{\lambda}{2}\int_0^t dt' \; \langle -| { J} e^{-(t-t')H}
\Delta H |\tilde \alpha \rangle
\end{equation}
Thanks to the previous established (\ref{bathresponse}), we have:
\begin{equation}
\Delta H  |\tilde \alpha \rangle =  - c_{\alpha}\; H' |\tilde \alpha \rangle
\label{cosa}
\end{equation}
where we have defined
\begin{equation}
\label{h1}
H'= [H_L - H_R, E].
\end{equation}
Substituting in (\ref{gk1}), we then have
\begin{equation}
\label{gk2}
\kappa_N =  \lim_{t\rightarrow\infty}
\,\frac{\lambda\,c_{\alpha}}{2}\;\int_0^t dt' \; \langle -| { J}
e^{-(t-t')H} H' |\tilde \alpha \rangle
\end{equation}

\subsection{Green-Kubo formula in the bulk}

The  Green-Kubo formula for the conductivity usually involves the
current temporal autocorrelation function \cite{TKS}. To see that formula
(\ref{gk2}) is indeed the standard Green-Kubo formula we proceed
further by using some  properties of the reservoirs.
Specifically, we use the assumption that the reservoir on the left
(resp. on the right) depends only from the phase/configuration
variables of the first (resp. last) site. This implies that
\begin{eqnarray}
\frac{1}{N} \,[A,H_L + H_R] & = & \frac{1}{N}\,\left[\sum_{s=1}^N sE_s
  -\frac{N+1}{2}\sum_{s=1}^N E_s \,,\, H_L+H_R\right] \nonumber\\ & = &
\frac{1}{N}\,\left[\left(\frac{1-N}{2}\right) E_1 \,,\, H_L\right] +
\frac{1}{N}\,\left[\frac{N-1}{2} E_N \,,\, H_R\right] \nonumber\\ & = &
\frac{N-1}{2N}\,[H_L-H_R,E]
\nonumber\\ & = & \frac{N-1}{2N}\, H'
\label{ggg}
\end{eqnarray}
On the other hand, we also have
\begin{equation}
\frac{1}{N} [A,H_R+H_L]= \frac{1}{\lambda N} [A,H-H_B]=
-\frac{1}{\lambda N} [H ,A] - \frac{J}{\lambda N}
\label{hhh}
\end{equation}
Putting (\ref{ggg}) and (\ref{hhh}) together we find
\begin{equation}
\frac{1}{2} H' = -\frac{1}{\lambda (N-1)} \left({ J} + [H ,A]\right)
\label{jjj}
\end{equation}
The standard Green-Kubo formula is obtained by substituting $H'$ on
the right of (\ref{gk2}) by the expression found in
Eq. (\ref{jjj}). We have
\begin{eqnarray}
\kappa_N & = & -\lim_{t\rightarrow\infty} \,
\frac{\,c_{\alpha}}{N-1}\;\int_0^t dt' \; \langle -|J
e^{-(t-t')H}J |\tilde \alpha \rangle \nonumber \\ & &
-\lim_{t\rightarrow\infty} \,
\frac{\,c_{\alpha}}{N-1}\;\int_0^t dt' \; \langle -|J
e^{-(t-t')H} [H,A] |\tilde \alpha \rangle
\end{eqnarray}
The term involving the commutator  $[H,A]$ can be further simplified as follows
\begin{eqnarray}
\lim_{t\rightarrow\infty} \int_0^t dt' \; \langle -|J
e^{-(t-t')H} HA |\tilde \alpha \rangle & = & \lim_{t\rightarrow\infty}
\int_0^t dt' \; \frac{d}{dt'} \langle -|J e^{-(t-t')H} A
|\tilde \alpha \rangle \nonumber\\ & = & \langle -|J A |\tilde \alpha
\rangle
\end{eqnarray}
where we have used that $H|\tilde \alpha\rangle = 0$ and we have assumed that
the $J$ and $A$ have vanishing correlation for very large
times.  Moreover, by using (\ref{maincommutator}) and the fact that
$H_B |\tilde \alpha \rangle = \langle - |H_B = 0$, one has that
\begin{eqnarray}
\langle -|JA |\tilde \alpha \rangle & = &
\langle -| [A,H_B]A |\tilde \alpha \rangle
\nonumber\\
&=&
\langle -| AH_BA |\tilde \alpha \rangle
\nonumber\\
&=&\frac{1}{2}\,\langle -| [[A,H_B],A] |\tilde \alpha \rangle\;.
\end{eqnarray}
Finally we  get
\begin{equation}
\label{gk3}
\kappa_N  =  -\lim_{t\rightarrow\infty} \,
\frac{\,c_{\alpha}}{N-1}\;\int_0^t dt' \; \langle -|J
e^{-(t-t')H}J |\tilde \alpha \rangle
-\, \frac{\,c_{\alpha}}{2(N-1)}\,\langle -|
[[A,H_B],A] |\tilde \alpha \rangle
\end{equation}
The second term comes from the fact, already mentioned above,
that $ \langle -|J e^{-(t-t')H}J |\tilde \alpha \rangle $ is,
 in stochastic systems,
 the current-current correlation function only up to an equal-time
extra term.
We distinguish two cases:
\begin{itemize}
\item $\lambda = 0$

In this case $H_{\lambda} = H_B$ and then the current autocorrelation function
can be treated as we just did for the extra term:

\begin{eqnarray}
\lim_{t\rightarrow\infty} \,\int_0^t dt' \; \langle -|J
e^{-(t-t')H_B}J |\tilde \alpha \rangle & = &
\lim_{t\rightarrow\infty} \int_0^t dt' \; \langle -|J
e^{-(t-t')H_B} H_B A |\tilde \alpha \rangle \nonumber \\ & = &
\lim_{t\rightarrow\infty} \int_0^t dt' \; \frac{d}{dt'} \langle -|
   J e^{-(t-t')H_B} A |\tilde \alpha \rangle \nonumber\\ & = &
    \langle -|J A |\tilde \alpha \rangle
    \nonumber\\ &=&\frac{1}{2}\,\langle -| [[A,H_B],A] |\tilde \alpha
    \rangle\;.
\end{eqnarray}
This implies that the two terms in Eq. (\ref{gk3}) cancel each others
and, consistently with the fact that there is no coupling to the
reservoirs, we find $\kappa_N = 0$ (see Ref. \cite{open} for a discussion)

\item $\lambda \neq 0$

This time $\kappa_N$ is different from zero and is, in general, the sum of
two competing terms: the time integral of the current autocorrelation function
and an extra-term which is purely due to randomness.

\vspace{0.5cm}
{\bf Example A:} {\em Hamiltonian system}

The commutator in Eq. (\ref{gk3}) vanishes in the Hamiltonian
case. Indeed we have:
\begin{eqnarray}
[A,H_B] f = A\{E,f\} - \{E,Af\} \qquad\qquad\forall f
\end{eqnarray}
from which it follows that
\begin{eqnarray}
[[A,H_B],A] f
&=&
[A,H_B] Af - A [A,H_B] f
\nonumber \\
& = &
A\{E,Af\} - \{E,A^2f\} - A^2\{E,f\} + A\{E,Af\}
\nonumber\\
& = &
0
\end{eqnarray}
Inserting $\alpha = T$ and $c_{T} = - \frac{1}{T^2}$ into (\ref{gk3}), the
Green-Kubo formula for the conductivity then reads
\begin{eqnarray}
\kappa_N & = & \lim_{t\rightarrow\infty} \,
\frac{1}{T^2(N-1)}\;\int_0^t dt' \; \langle -|J e^{-(t-t')H} J |\beta \rangle
\end{eqnarray}

\vspace{0.5cm}
{\bf Example B:} {\em Simple symmetric exclusion process}

In this case the commutator in Eq. (\ref{gk3}) is non-zero. We have
\begin{equation}
[A,H_B] =J = \sum_{s=1}^{N-1} \left(S_s^{-}S_{s+1}^{+} -
S_s^{+}S_{s+1}^{-}\right)
\end{equation}
and
\begin{equation}
[[A,H_B],A] = -\sum_{s=1}^{N-1} \left(S_s^{-}S_{s+1}^{+} +
S_s^{+}S_{s+1}^{-}\right)
\end{equation}
Recalling that for the SSEP with reservoirs having the same chemical
potential the equilibrium state is given by a Bernoulli product
measure, see Eq. (\ref{sep-eq}), an immediate computation gives
\begin{eqnarray}
\langle -| [[A,H_B],A] |\tilde \alpha \rangle
& = &
-2\rho(1-\rho)(N-1)
\end{eqnarray}
This implies the following expression for the conductivity:
\begin{eqnarray}
\label{okok}
\kappa_N & = & -1 + \lim_{t\rightarrow\infty}
\, \frac{1}{(N-1)\rho(1-\rho)}\;\int_0^t dt' \; \langle -|J
e^{-(t-t')H}J |\rho \rangle
\end{eqnarray}

As a final remark of this section, let us check that
in the thermodynamical limit the correct value of the
conductivity is recovered. In order to evaluate the contribution due to the current
autocorrelation function we observe that in the thermodynamic limit
$N\rightarrow\infty$, whatever the value of $\lambda < \infty$, the
boundaries will be negligible.  This lead us to evaluate this term for
the {\em infinite volume system}:
\begin{eqnarray}
\label{infinite}
& & \lim_{N\rightarrow\infty}\lim_{t\rightarrow\infty} \,
\frac{1}{(N-1)\rho(1-\rho)}\;\int_0^t dt' \; \langle -|J
e^{-(t-t')H}J |\rho \rangle = \nonumber\\ & & =
\frac{2}{\rho(1-\rho)}\;\int_0^{\infty} dt' \; < {\cal J}_{0,1}(0)
{\cal J}_{0,1}(t') > \nonumber \\
\end{eqnarray}
where $<\cdot >$ denotes expectation with respect to the equilibrium state.
We can then use the duality property for the model and specifically
the following results:
\begin{equation}
<\left(n_{x} (t) - \rho\right)\left(n_{0} (0) - \rho\right)> \; = \;
\rho (1-\rho)\, p_t(x)
\end{equation}
where $p_t(x)$ is the probability that a continuous time simple
symmetric random walk jumping left or right at rate 1, started at the
origin at time zero, is found at site $x$ at time $t$.  Using duality
we have
\begin{eqnarray}
\label{dual}
< {\cal J}_{0,1}(0) {\cal J}_{0,1}(t') >
& = &
< \left(n_1(0) - n_0(0) \right) \left(n_{1}(t') - n_{0}(t') \right) >
\nonumber\\
& = & \rho(1-\rho) \left(  - p_{t'}(-1) + 2p_{t'}(0) - p_{t}(1) \right)
\nonumber\\
& = & - \rho(1-\rho) \frac{d}{dt'} p_{t'}(0)
\end{eqnarray}
Putting together (\ref{infinite}) and (\ref{dual}) we arrive to
\begin{eqnarray}
\lim_{N\rightarrow\infty}\lim_{t\rightarrow\infty} \,
\frac{1}{(N-1)\rho(1-\rho)}\;\int_0^t dt' \; \langle -|J
e^{-(t-t')H}J |\rho \rangle &=& -2(p_{\infty}(0) - p_{0}(0))
\nonumber\\ &= & 2
\end{eqnarray}
Inserting this result in Eq. (\ref{okok}) we finally find (as it should be!):
\begin{equation}
k = \lim_{N\rightarrow\infty} \kappa_N = -1 + 2 = 1
\end{equation}
\end{itemize}










\subsection{Green-Kubo formula for the boundaries}

Here we follow the opposite strategy of the previous section.
Recall expression (\ref{gk2}) for the conductivity:
\begin{equation}
\kappa_N = \lim_{t\rightarrow\infty} \,
\frac{\lambda\,c_{\alpha}}{2}\;\int_0^t dt' \; \langle -|J
e^{-(t-t')H} H' |\tilde \alpha \rangle
\end{equation}
This time we express ${ J}$ in terms of $H'$ by inverting relation
(\ref{jjj}), that is:
\begin{equation}
{ J} = -\frac{\lambda (N-1)}{2} H' - [H,A]
\end{equation}
This yields the following expression
\begin{eqnarray}
\kappa_N & = & -\lim_{t\rightarrow\infty} \,
\frac{\lambda^2\,c_{\alpha}(N-1)}{4}\;\int_0^t dt' \; \langle -| H'
e^{-(t-t')H} H' |\tilde \alpha \rangle \nonumber \\ & &
-\lim_{t\rightarrow\infty} \,
\frac{\lambda\,c_{\alpha}}{2}\;\int_0^t dt' \; \langle -| [H,A]
e^{-(t-t')H} H' |\tilde \alpha \rangle
\end{eqnarray}
As in the previous case the term involving the commutator $[H,A]$ can
be simplified by using the fact that it is the integral of a
time-derivative, and that the correlations of $ A$ and $ H'$ vanish at
widely separated times:
\begin{eqnarray}
\kappa_N & = & -\lim_{t\rightarrow\infty} \,
\frac{\lambda^2\,c_{\alpha}(N-1)}{4}\;\int_0^t dt' \; \langle -| H'
e^{-(t-t')H} H' |\tilde \alpha \rangle \nonumber \\ & &
- \, \frac{\lambda\,c_{\alpha}}{2}\;\langle
-| AH' |\tilde \alpha \rangle
\end{eqnarray}
The extra term can be rearranged as follows. Recalling the definition
of $H'$, Eq. (\ref{h1}), and using the fact that $H_L$ and $H_R$
annihilate the equilibrium measure we have
\begin{eqnarray}
\langle -| AH' |\tilde \alpha \rangle
& = &
\langle -| A[H_L-H_R,E] |\tilde \alpha \rangle
\nonumber \\
& = &
\langle -| A(H_L-H_R)E |\tilde \alpha \rangle
\end{eqnarray}
Now we use that $H_L$ (resp. $H_R$) commutes with all the $E_i$ with
$i \neq 1$ (resp. $i\neq N$). This yields:
\begin{eqnarray}
\langle -| AH' |\tilde \alpha \rangle
& = &
- \frac{N-1}{2} \langle -| E (H_L + H_R) E] |\tilde \alpha \rangle
\nonumber \\
& = &
\frac{N-1}{4}  \langle -| [E,[E, H_L + H_R]] |\tilde \alpha \rangle
\end{eqnarray}
and the final expression for the Green-Kubo formula for the boundaries read:
\begin{eqnarray}
\kappa_N & = & -\lim_{t\rightarrow\infty} \,
\frac{\lambda^2\,c_{\alpha}(N-1)}{4}\;\int_0^t dt' \; \langle -| H'
e^{-(t-t')H} H' |\tilde \alpha \rangle \nonumber \\ & &
-\, \frac{\lambda\,c_{\alpha}(N-1)}{8}\;\langle
-| [E,[E, H_L + H_R]] |\tilde \alpha \rangle
\label{condu}
\end{eqnarray}
We will show below that the extra term is non-zero for both the
Hamiltonian chain and the SEP and has actually the same value.

\vspace{0.5cm}
{\bf Example A:} {\em Hamiltonian system}

The extra term is evaluated as
\begin{eqnarray}
\langle -| [E,[E, H_L + H_R]] |\beta \rangle & = & \langle -|
        [E_1,[E_1, H_L]] |\beta \rangle + \langle -| [E_N,[E_N, H_R]]
        |\beta \rangle \nonumber\\ & = & -2 \langle -| E_1H_L E_1
        |\beta \rangle - 2 \langle -| [E_N H_R E_N |\beta \rangle
\end{eqnarray}
We have
\begin{eqnarray}
\langle -| E_1H_L E_1 |\beta \rangle & = & -\langle -|
\frac{p_1^2}{2}\frac{\partial}{\partial p_1}\left(T
\frac{\partial}{\partial p_1} + p_1\right)\frac{p_1^2}{2} |\beta
\rangle \nonumber\\ & = & T \langle -| p_1^2 |\beta \rangle
\nonumber\\ & = & T^2
\end{eqnarray}
where we have used the fact that $\frac{\partial}{\partial p_1}$
annihilates the flat measure to the left and $\left(T
\frac{\partial}{\partial p_1} + p_1\right)$ annihilates the
equilibrium measure to the right, together with
\begin{eqnarray}
\left[\frac{p_1^2}{2}, \frac{\partial}{\partial p_1}\right] = - p_1
\end{eqnarray}
\begin{eqnarray}
\left[T \frac{\partial}{\partial p_1} + p_1, \frac{p_1^2}{2}\right] = T p_1
\end{eqnarray}
Analogously we find that
\begin{eqnarray}
\langle -| E_NH_R E_N |\beta \rangle
& = &
T^2
\end{eqnarray}
so that the final result for the extra term in the thermodynamic limit is:
\begin{equation}
\frac{\lambda\,c_{\alpha}(N-1)}{8}\;\langle -| [E,[E, H_L + H_R]] |\beta \rangle =
\frac{\lambda (N-1)}{2}
\end{equation}

\vspace{0.5cm}
{\bf Example B:} {\em Simple symmetric exclusion process}

With
\begin{equation}
H_L = -\rho\left(S_1^{+} + S_1^{0} - \frac12 \right) -
(1-\rho)\left(S_1^{-} - S_1^{0} - \frac12 \right)
\end{equation}
we have
\begin{eqnarray}
\langle -| [E,[E, H_L]] |\rho \rangle
& = &
\langle -| [E_1,[E_1, H_L]] |\rho \rangle
\nonumber\\
& = &
\langle -| -\rho S_1^+ - (1-\rho) S_1^-|\rho \rangle
\end{eqnarray}
Because $(S_1^+ + S_1^0-\frac12)$ and $(S_1^- - S_1^0-\frac12)$
separately annihilate the flat measure, we can substitute $S_1^{\pm}$ by
$S_1^{0}$'s to get
\begin{eqnarray}
\langle -| [E,[E, H_L]] |\rho \rangle
& = &
\langle -| \rho \left(S_1^0 -\frac12\right) + (1-\rho) \left( -S_1^0 - \frac12\right) |\rho\rangle \nonumber\\
& = &
-\rho(1-\rho) - (1-\rho) \rho\nonumber\\
& = &
-2\rho(1-\rho)
\end{eqnarray}
The extra term is then:
\begin{equation}
\frac{\lambda\,c_{\alpha}(N-1)}{8}\;\langle -| [E,[E, H_L + H_R]] |\rho \rangle =
\frac{\lambda\,(N-1)}{2}
\end{equation}

\section{Stiffness}

Let us now compute  the stiffness.  We have a
`quantum-like' Hamiltonian $H=H_B+\lambda(H_L+H_R)$ where the bulk
term is ``symmetric'' with respect to transformations generated by
$E$, since $[H_B,E]=0$.  $H_L$ and $H_R$ are the boundary `handles'
that break the symmetry generated by $E$. In order to calculate the
stiffness, we {\bf twist} the boundaries, i.e. we apply the
transformation in opposite directions:
\begin{equation}
H_L \rightarrow  H_L^\theta = e^{i\theta E} H_L  e^{-i\theta E} \qquad\qquad
H_R \rightarrow  H_R^\theta = e^{-i\theta E} H_R  e^{i\theta E}
\end{equation}
As a consequence of the twist the spectrum of the transformed
Hamiltonian $ H^\theta = H_B + \lambda ( H_L^\theta + H_R^\theta)$ will
be different from the one of the original Hamiltonian $H$.
We define the stiffness $\sigma$ in terms of
 the lowest eigenvalue of $ H^\theta$:
\begin{eqnarray}
\epsilon^\theta & = &
-  \lim_{t\rightarrow\infty}
\frac{\ln\left(\langle-| e^{-t H^\theta}|\psi(t)\rangle\right)}{t}
\end{eqnarray}
We define
\begin{equation}
  H^\theta = H_L^\theta +  H_R^\theta
\end{equation}
Developing up to order $\theta^2$ we have
\begin{eqnarray}
\Delta H & \equiv  & H^\theta-H =e^{i\theta E} H_L e^{-i\theta E} + e^{-i\theta E}
H_R e^{i\theta E} - (H_L-H_R) \nonumber \\
&
= &  - i\theta H' - \frac{\theta^2}{2}
\left([E,[E,H_L+H_R]]\right) +\ldots \nonumber
\end{eqnarray}
Application of time-dependent perturbation theory gives
\begin{eqnarray}
\ln\left(\langle-| e^{-t\tilde H}|\psi(t)\rangle\right) & = &
\ln\left(1 + \lambda \int_0^t dt' \langle-|\Delta \tilde H
|\tilde \alpha\rangle\right.  \\ & & \quad\quad\; + \left.\frac{\lambda^2}{2}
\int_{0}^{t} dt' \int_{0}^t dt'' \langle-|\Delta \tilde H
e^{-(t''-t')H_B}\Delta \tilde H |\tilde \alpha\rangle +\ldots \right)
\nonumber\\ & = & \ln\left(1 - \theta^2\;\frac{\lambda}{2} \;t\;
\langle-| \left([E,[E,H_L+H_R]]\right) |\tilde \alpha\rangle\right.
\nonumber\\ & & \quad\quad - \left.\theta^2\frac{\lambda^2}{2}
\int_{0}^{t} dt' \int_{0}^t dt'' \langle-| H' e^{-(t''-t')H_B}H'
|\tilde \alpha\rangle +\ldots \right) \nonumber
\end{eqnarray}
We find for the increase in the lowest eigenvalue
\begin{eqnarray}
\epsilon^\theta -\epsilon^o
& \sim & \frac{1}{2}  \frac{\sigma_N}{(N-1)} (2\theta)^2
\nonumber\\
& = &
-  \lim_{t\rightarrow\infty}
\frac{\ln\left(\langle-| e^{-t\tilde H}|\psi(t)\rangle\right)}{t}
\nonumber\\
& = &
-  \lim_{t\rightarrow\infty}
\frac{\theta^2\lambda}{2} \;\langle-|
\left([E,[E,H_L+H_R]]\right)  |\tilde \alpha\rangle
\nonumber\\
& &
-  \lim_{t\rightarrow\infty}
\theta^2\lambda^2\int_{0}^{t} dt'
\langle-| H' e^{-(t-t')H_B}H' |\tilde \alpha\rangle
\nonumber
\end{eqnarray}
Recalling the expression (\ref{condu}) for the conductivity
\begin{eqnarray}
\frac{2 \kappa_N}{c_{\alpha}(N-1)} & = &
-\lim_{t\rightarrow\infty} \, \frac{\theta^2\lambda^2}{2}\;\int_0^t dt' \;
\langle -| H' e^{-(t-t')H} H' |\tilde \alpha \rangle \nonumber \\ & &
-\lim_{t\rightarrow\infty} \, \frac{\theta^2\lambda}{4}\;\langle -|
[E,[E, H_L + H_R]] |\tilde \alpha \rangle
\end{eqnarray}
one obtains the following relation between the conductivity
and the stiffness:
\begin{equation}
\kappa_N  = c_{\alpha} \sigma_N\;.
\end{equation}
Stiffness and conductivity are proportional up to a trivial factor.

\section{Conclusions}

When a system has a conserved bulk quantity, its bulk evolution operator
has a symmetry. Current transport
 induced by boundary `reservoir'
terms corresponds in all generality to a twist exerted
 applying the symmetry transformation in opposite senses
to the  boundary terms. The conductivity of the system is in this view
the stiffness, or helicity modulus, associated with  this operation.

This mechanical analogy of transport can be taken further. For example,
one easily understands that in an elastic system
 any local perturbation that couples with torsion will have long-range effects:
 the analogy discussed here means that the same can be said of
perturbations  of systems with conserved quantities \cite{kirkpatrick}.

In fact, one recognizes methods where auxiliary thermal baths are used,
with their temperature fixed so that they exchange no current on
average~\cite{Visscher} as the usual symmetry-breaking fields, adjusted
so that they exert no average force, of statistical mechanics.

The derivation in this paper was made for classical and stochastic systems,
in contact with reservoirs at the ends. The generalization to a quantum system
with baths can be made through the Feyman-Vernon~\cite{cl} formalism.
It is not at this point clear to us how this may be related
to the approach of Kohn~\cite{Kohn} and Shastry and Sutherland~\cite{Shastry},
where the twist is applied to the Hamiltonian
(rather than the evolution operator) of a closed chain.

\end{document}